\documentclass[12pt]{article}
\usepackage{jcappub}
\usepackage[english]{babel}
\pdfoutput=1
\usepackage[LGR,T1]{fontenc}
\usepackage[utf8,applemac,latin1]{inputenc}
\usepackage{color,endnotes}
\usepackage{hyperref}
\usepackage{xcolor}
\usepackage{multirow}
\usepackage{aas_macros}
\usepackage[nottoc]{tocbibind}

\begin{document}
\title{Combining strong and weak lensing estimates in the \textsc{Cosmos} field}


\author[a,b]{Felix Arjun Kuhn}
\author[c]{, Simon Birrer$^*$}
\author[a]{, Claudio Bruderer}
\author[a,d]{, Adam Amara}
\author[a]{and Alexandre Refregier}

\affiliation[a]{Institute for Particle- and Astrophysics, ETH Zurich, Wolfgang-Pauli-Strasse 27, 8093, Zurich, Switzerland}
\affiliation[b]{Institute for Theoretical Physics, Heidelberg University, Philosophenweg 12, 69120, Heidelberg, Germany}
\affiliation[c]{Kavli Institute for Particle Astrophysics and Cosmology and Department of Physics, Stanford University, Stanford, CA 94305, USA}
\affiliation[d]{Institute of Cosmology \& Gravitation, University of Portsmouth, Dennis Sciama Building, Burnaby Road, Portsmouth PO1 3FX, UK}

\emailAdd{sibirrer@stanford.edu}
\emailAdd{felix.kuhn@stud.uni-heidelberg.de}

\abstract{
We present a combined cosmic shear analysis of the modeling of line-of-sight distortions on strongly lensed extended arcs and galaxy shape measurements in the COSMOS field.
We develop a framework to predict the covariance of strong lensing and galaxy shape measurements of cosmic shear on the basis of the small scale matter power-spectrum.
The weak lensing measurement is performed using data from the COSMOS survey calibrated with a cloning scheme using the Ultra Fast Image Generator \textsc{UFig} \citep{berge2013}.
The strong lensing analysis is performed by forward modeling the lensing arcs with a main lensing deflector and external shear components from the same \textit{Hubble Space Telescope} imaging data set.
With a sample of three strong lensing shear measurements we present a 2-sigma detection of the cross-correlation signal between the two complementary measurements of cosmic shear along the identical line of sight. With large samples of lenses available with the next generation ground and space based observatories, the covariance of the signal of the two probes with large samples of lenses allows for systematic checks, cross-calibration of either of the two measurement and the measurement of the small scale shear power-spectrum.
}
\arxivnumber{2010.08680}

\maketitle

\section{Introduction}


Gravitational lensing is a unique phenomenon, as it allows to probe the distribution of matter and its evolution through cosmic time within the Universe at a wide range of scales.
This phenomenon is typically analysed depending on the effects' strength, where one broadly distinguishes between weak and strong gravitational lensing.\\
Weak lensing describes the linear distortion, primarily of galaxy shapes or the cosmic microwave background temperature and polarization patterns, due to the large scale structure of the Universe. In wide-field galaxy surveys, the lensing signal is measured by correlating a large number of observed galaxy shapes.\\
Weak lensing has provided competitive constraints on the dark matter content in the universe and clustering strength of large scale structure.\\
\citep[e.g.][]{hikage2019, hildebrandt, troxel, asgari2020}. 
There is an emergent discrepancy between the weak lensing clustering signal and the CMB evolved expectation to late times, observed by independent survey teams.\\
As precision has increased over the last decade, investigating remaining systematic uncertainties in the weak lensing measurement has become a major focus of current analyses. Of concern are, among others, effects associated with intrinsic alignments, photometric redshift estimates, blending effects, and the calibration of the shape noise removal \citep[e.g.][]{mandelbaum2018,ghosh2020}. \\
Strong lensing, on the other hand, describes the extreme deflection of light by massive objects, like individual galaxies, along the line of sight, which lead to multiple images of the very same source. The uniqueness of multiple images of the same source allows one to make precise measurements of lensing quantities (such as Einstein radius, ellipticity and external shear) without relying on priors on the shape of the source at high precision per individual lens.
Strong lenses have been utilized to provide strong constraints on dark matter physics \citep[e.g.][]{hezaveh2016, birrer_wdm, vegetti2018, hsueh2020, gilman20}, the mass density profiles and evolutionary paths of massive elliptical galaxies \citep[e.g.][]{sonnenfeld2013, shajib2020_slacs} and the Hubble constant $H_0$
\citep[e.g.][]{wong20, shajib2020, millon2020, birrer2020}
These current results have been made with a comparably small number of strong lenses (of order 10).\\
The leading systematic uncertainties in current analyses of strong lenses arises from uncertainties in the mass density profiles of the main deflector galaxies \citep[see e.g.][for its implication and inference incorporating these uncertainties]{birrer2020} and potential selection effects in the lensing efficiency, originating from both the main deflector and line-of-sight structure.\\
In many regards, strong and weak lensing measurements are driven by the same science questions, competitive in their achieved precision as stand-alone probes. They are mostly independent and complementary in their approaches and systematic uncertainties and can shed light on the existing and emerging tensions within the $\Lambda CDM$ cosmological model. \\
Inherently, strong and weak lensing probe the identical matter density field and thus their measurements share covariances. Their cross-correlation signal can add valuable information to the cosmological signal and, likewise, be used as a diagnostic for unaccounted systematics in either of the two probes. Weak lensing measurements have thus been used to inform strong lensing measurements by adding constraints on the line-of-sight convergence at a strong lens position \citep{Tihhonova:2018, Tihhonova:2020}.\\
On the other hand, strong lensing observables inherently also contain the line-of-sight weak shear distortions.
In current strong lensing analyses, linear external shear components are an integral part of the lens model, and thus provide information about the large scale structure of the universe themselves.
\cite{birrer-welschen} investigated the non-linear shear distortions being imprinted in the shape of Einstein rings and \cite{birrer-rings} proposed to use strong lensing shear information as a complementary probe of cosmic shear to galaxy shape weak lensing.
To this date, no joint analysis of the clustering signal from strong and weak lensing observables has been performed.
A pre-requisit of a combined analysis is exquisite analyses on both sides (weak and strong lensing) as well as a self-consistent formalism in describing the measurements and their covariances when comparing them to an underlying theory prediction.
In this work, we perform both a strong and a weak lensing analysis in the COSMOS field using imaging data by the \textit{Hubble Space Telescope} (HST). We develop a self-consistent theoretical framework derived from the matter power spectrum to compare both observables. The $\sim$ 2 deg$^2$ field is a pilot project that can readily be further scaled up in area with the upcoming ground- and space-based surveys (LSST\footnote{https://www.lsst.org/}, Euclid\footnote{https://www.euclid-ec.org/}, Roman Space Telescope\footnote{https://roman.gsfc.nasa.gov/}). The goal of this paper is to assess the potential of forthcoming larger scale joint analyses and to investigate systematics of the current approaches to measure the weak and strong lensing signals.
In this paper, we extract the lensing information in the strong lensing regime by performing the lens reconstruction with basis sets as described in \cite{lenstronomy, birrer-basis}. To estimate the weak lensing signal from the shape measurements of galaxies, we employ ultra-fast image simulations by \textsc{UFig} \citep{berge2013} to emulate the COSMOS images and calibrate the weak lensing shape measurement. We apply these methods and compare the estimated shear values at the positions of the strong lens systems. 
We find that with the strong lensing samples expected with the upcoming ground and space-based wide field surveys, there will be  sufficient statistical information in the covariance signal to impact a joint analysis. On one hand, the expected signal can enable the calibration of weak lensing shear measurement and redshift distribution uncertainties with strong lensing measurement. On the other hand
The conducted analysis in this work serves as a exploratory study which enables both avenues.\\
The paper is structured as follow:
In Section \ref{sec:theory} we elaborate on the theory of gravitational lensing and our self-consistent formalism to treat the lensing effect in different regimes. In Section \ref{sec:data} we introduce the COSMOS dataset we work with. In Section \ref{sec:sl-analysis} we discuss the analysis of three strong lens systems from this dataset with the method by \citep{birrer-basis}. We use the same data for a weak lensing analysis in Section \ref{sec:wlanalysis}, making use of the cloning-method based on \citep{bruderer, cb2016b, cb2018}. The strong and weak lensing measurement results, including the computation of the covariances of the observables, are compared in Section \ref{sec:joint-analysis}. We conclude in Section \ref{sec:discussion}.

\section{Gravitational lensing in different regimes}
\label{sec:theory}
Gravitational lensing describes the deflection of light by intervening over- or under-dense structures along the line of sight as predicted and described by General Relativity \citep{mb2017,ar2003}.
Typically, gravitational lensing phenomena are characterized by the strength of the deflection and two distinct regimes are considered: {\it Strong} lensing describes the effect of strongly distorted and multiplied images of the very same source and is typically caused by a massive object (i.e. a single, massive galaxy or a galaxy cluster) in between the observer and the source. {\it Weak} lensing describes the small deformations of the objects' visible shapes due to massive structures along the line of sight towards the observer. It occurs in the absence of a rare chance-alignment of a massive dominant lensing structure acting as a lens and the source and is thus the more prevalent effect. \\
In this section, we review the theory and formalism of gravitational lensing in order to develop a consistent framework of the phenomenon for both strong and weak lensing observables. We use the conventions presented in \citep{birrer-welschen} and \citep{mb2010}.\\
We describe the formalism for describing weak lensing shear distortions (\ref{subsec:shear}), the modifications needed to describe strong lensing shear (\ref{subsec:observables}), the method to calculate power spectra (\ref{subsec:shear-ps}) and cross-correlated power spectra (\ref{subsec:wl-sl-covariance}). 

\subsection{Weak lensing shear distortions}
\label{subsec:shear}
The lowest order distortions of the galaxy shapes due to weak gravitational lensing are called shear. Shear describes the uniform elliptical distortion of an image (i.e. a perfect circle would be deflected and seen as an ellipse) and is the first order term of weak lensing image distortions. Shear is defined component-wise as double derivative derivatives of the so-called lensing potential 
\begin{equation}
\label{eq:lensingpot}
    \psi(\theta^j) = 2 \int_0^{w_s} \textup{d}w' \frac{w_s-w'}{w' w_s} \phi \left(w'\theta^j,w'\right),
\end{equation}
\begin{equation}
\label{eq:pot-shear-rel}
    \begin{aligned}
    &\gamma_1 = \frac{1}{2}(\partial^1 \partial_1 \psi - \partial^2 \partial_2 \psi), \\
    &\gamma_2 = \partial^1 \partial_2 \psi = \partial^2 \partial_1 \psi,
    \end{aligned}
\end{equation}
the shear $\gamma$ is then defined as $\gamma = \gamma_1 + i\gamma_2$. \\
For the measurement of the weak lensing shape deformations, one has to measure the (lensed) shapes of a multitude of galaxies and correlate them, as their intrinsic (unlensed) shapes are unknown. For this work, we assume that the intrinsic galaxy ellipticities are, to leading order, uncorrelated and that they are drawn from distributions at mean zero. Thus, the arithmetic average over the intrinsic ellipticities of an ensemble of N galaxies is assumed to be zero. The measured shape is then a composite of the intrinsic source ellipticity and the deformation induced by cosmic shear
\begin{equation}
    \varepsilon = \varepsilon^i + \gamma,
\end{equation} where $\varepsilon$ is the measured ellipticity as observed on the sky, $\varepsilon^i$ is the intrinsic ellipticity and $\gamma$ the cosmic shear. By averaging over the measured shape of the whole ensemble the intrinsic shapes averages to zero, $\left\langle \varepsilon^i \right\rangle_N = 0$. Here, $\left\langle A \right\rangle_N$ denotes the statistical average of the observable A over an ensemble of N galaxies. The remainder of the measured mean ellipticities can be identified as the mean cosmic shear of the analysed region
\begin{equation}
    \left\langle\varepsilon \right\rangle_N = \left\langle \varepsilon^i \right\rangle_N  + \left\langle\gamma \right\rangle_N  = \left\langle\gamma \right\rangle_N
    .
\end{equation}
The shear susceptibility is then introduced in order to relate the averaged ellipticity to the gravitational shear: 
\begin{equation}
\label{eq:shear-susc}
\left\langle \gamma \right\rangle = G^{\gamma} \left\langle \gamma \right\rangle_N .
\end{equation}
The measurements of the galaxy shapes as well as of the parameter $G^{\gamma}$ are challenging tasks. We describe our methods to tackle those in section \ref{sec:wl-analysis}.\\
While weak lensing yields estimates of the average cosmic shear in regions of space, the cosmic shear at special, specific points can be estimated by using strong lensing \citep{birrer-welschen}. This is described in more detail in Section \ref{sec:sl-analysis}. We note that these yield independent estimates of the same underlying quantity. While similar, the estimates have subtle differences we elaborate on in the following sections. Understanding these is necessary for a careful treatment of the systematics of each of the measurements.

\subsection{Strong and weak lensing kernels} \label{subsec:observables}
By considering the expression for the lensing (deflection) potential as given in equation \ref{eq:lensingpot}, we note that the expression for the lensing potential needs to be modified for a strong lensing system. One can consider the kernel function of the weak lensing, 
\begin{equation}
    n(w)^{WL} = \frac{w_s - w}{w_s w},
\end{equation}
 a continuous smoothing function depending on the galaxies averaged over for the weak lensing measurement. The kernel function of the strong lensing is not a smooth function because the deflector is at a single position while for weak lensing there are multiple deflectors at different positions. The strong lensing kernel must therefore contain a sharp delta-function, singling out the (comoving) distance $w_s$ of the lensing galaxy 
\begin{equation}
    n(w)^{SL} = \delta (w - w_s).
\end{equation}
This also implies that the real space shear, as measured by the weak lensing analysis, is a smoothed function, strongly depending on the selected galaxies.

\subsection{Shear power spectra}
\label{subsec:shear-ps}
In order for the cosmic shear estimates by weak and strong lensing to be correlated, as described in the next section, we need to derive the angular shear power spectra.\\
The power spectrum of the Newtonian gravitational potential $\phi$ which is assumed to be weak in this context, $\phi \ll c^2$, is defined as \citep[cf.][]{mb2010}
\begin{equation}
    \left\langle \hat{\phi}(q) \hat{\phi}^*(q') \right\rangle \equiv (2 \pi)^3 \delta_D(q-q') P_\phi(q)
\end{equation}
where $\phi$ was expanded into Fourier modes $\phi(x) = \int \frac{\textup{d}^3q}{ (2 \pi)^3} \hat{\phi}(q) e^{iqx}$. By computing the variance of the lensing potential (see equation \ref{eq:lensingpot}), we find the angular correlation function for the lensing potential 
\begin{equation}
    \left\langle \hat{\psi}(\theta) \hat{\psi}(\theta') \right\rangle = \int_0^{w_s} \textup{d}w  \int_0^{w_s} \textup{d}w' \frac{w_s-w}{w_s w} \frac{w_s-w'}{w_s w'} \int \frac{\textup{d}^3 q}{(2 \pi)^3} P_\phi (q,w) e^{iq(x-x')},
\end{equation}
where $x = (w \theta, w)$ and $\theta$ denotes the observation angle. Since the observations are made not on flat space but on the sky, i.e. a sphere, we have to introduce coordinates applicable to spherical surfaces. Therefore, we expand the Fourier basis into spherical harmonics $Y_{lm}$ as well as the effective lensing potential $\psi$, leading to expressions $\psi_{lm}$ with $\psi(\theta) = \sum_{lm} \psi_{lm} Y_{lm}$. These in turn define the angular lensing power spectrum $C_l$ by $\left\langle \psi_{lm} \psi_{l'm'}^* \right\rangle = \delta_{ll'} \delta_{mm'} C_l^\psi$. Looking at the large scale structure, one can further simplify the result by using the Poisson equation 
\begin{equation}
        C_l^\gamma = \frac{9}{4} \left( \frac{H_0}{c} \right)^4 \Omega^2_{m0}  \int_0^{w_s} \textup{d}w \left( \frac{w_s-w}{w_s a(w)}\right)^2 P_\delta   \left( \frac{l}{w}\right),
\end{equation}
which is a angular weak lensing power spectrum. 
To then estimate the variance of the shear signal for lensing, we evaluate 
\begin{equation}
\label{eq:shearvar}
\sigma^2_\gamma = \frac{1}{2 \pi} \int\displaylimits_{\ell_{min}}^{\ell_{max}} \left| C_\ell ^\gamma \right| \ell (\ell+1)  \mathrm{d} \log \ell,
\end{equation}
where $\log \ell$ denotes the natural logarithm of $\ell$ and $\ell_{min}$ and $\ell_{max}$ are set by the maximum and minimum angular scales of our analysis. With a slightly adapted expression for the angular power spectrum $C_l$ this equation is applicable for the calculation of variances for strong lensing as well.\\
Since we combine weak and strong lensing anlyses, the minimum angular scale is set by the length scale of strong lens systems ($\lesssim 1$ arcsec), which we approximate by setting $\ell_{max} \rightarrow \infty$. \\

\subsection{Combining strong and weak lensing}
\label{subsec:wl-sl-covariance}
To correlate the strong and weak lensing shear estimates, we define the shear measures
\begin{align}
    {\gamma}_{SL} &:= \frac{1}{A_{SL}} \int_{A_{SL}} \gamma_{z=z_{SL}} (x) \textup{d}x, \\
    {\gamma}_{WL} &:= \frac{1}{A_{WL}} \int_{A_{WL}} \int_z n(z) \gamma_{z} (x) \textup{d}x \textup{d}z,
\end{align}
where from here on $n(z)$ denotes the weak lensing kernel $n^{WL}(z)$. Thus, we average the shear over the respective considered regions (see Section \ref{subsec:observables}).  Shear is a combination of different second-order derivatives of the lensing potential $\psi$ (see equation \ref{eq:pot-shear-rel}).
By using the notation $O^{ij}(x) = \partial^i \partial^j O(x)$, one can then derive 
\begin{equation}
\psi_{SL}=\frac{1}{A_{SL}} \int_{A_{SL}} \int_w \textup{d}w \textup{d}x f_{d,s}(w) \frac{w_s-w}{w_s w} \phi(w,x),
\end{equation} 
\begin{equation}
\psi_{WL}=\frac{1}{A_{WL}} \int_{A_{WL}} \int_z \int_w \textup{d}w \textup{d}x \textup{d}z n(z) \frac{w(z)-w}{w(z) w} \phi(w,x),
\end{equation}
where we have introduced $f_{d,s}(w) = \left( 1- \frac{D_{dk}(w)}{D_k(w)} \frac{D_s}{D_{ds}(w)} \right)$ as the weight function in the strong lensing potential, which accounts for the light perturbation at the deflection slice $k$.
The covariance of the second-order derivatives of the potentials of the different lensing regimes is then
\begin{align}
     \left\langle \psi_{SL}^{ij} \psi_{WL}^{ij} \right\rangle = \int_{z'} \int_w f_{d,s}(w) n(z') \frac{w_s-w}{w_sw}\frac{w(z')-w}{w(z')w} \int_{k_{min}}^{k_{max}} P_\phi (k, w) dw dz dk.
\end{align}
The cross-correlation power spectrum is thus
\begin{equation}
\label{eq:cls-offdiag}
    C_{l_{SL} l_{WL}}(k):=  \int_{z'} \int_w f_{d,s}(w) n(z') \frac{w_s-w}{w_sw}\frac{w(z')-w}{w(z')w} P_\phi (k, w) dw dz,
\end{equation}
In this derivation, we have applied Born approximation \citep[][]{birrer-welschen}. Finally, the cross-correlation function of the lensing potentials is
\begin{equation}
    \left\langle \psi_{SL}^{ij} \psi_{WL}^{ij} \right\rangle = \partial^i \partial^j \int_{k_{min}}^{k_{max}} C_{l_{SL}l_{WL}}(k) \textup{d}k 
\end{equation}
which is a method of correlating strong and weak lensing estimates at different scales. Lastly according to eq. \ref{eq:pot-shear-rel} then, the left-hand side of this equation corresponds to the shear cross-correlation function $\left\langle \gamma_{SL} \gamma_{WL} \right\rangle$.

\section{Data}
\label{sec:data}
For our measurement of the cross-correlation between cosmic shear as measured with galaxy shapes (weak lensing) and with strong lensing, we use data from the COSMOS\footnote{http://cosmos.astro.caltech.edu/} survey \citep{cosmos}. 
The COSMOS field covers an area of approximately 2 square degrees centered at RA=10:00:28.6, DEC=+02:12:21.0 (J2000). The center of the region was covered with a multi-band imaging survey. The field was imaged with the \emph{Advanced Camera for Surveys} (ACS), which is mounted on the Hubble Space Telescope, during 590 orbits. Additionally, the area was also covered by spectroscopic surveys.
The field contains about one million identified galaxies \citep{leauthaud}. Their properties were then further measured using different telescopes. For instance, photometric redshifts for the most recent catalog have been estimated with ground-based telescopes using data from 26 different filters \citep{laigle}.  \\
For both the strong lensing and the weak lensing analysis, we work with the rotated and \emph{drizzled} v2 images from the F814W filter \citep{koekemoer}. The median exposure time of the images is $4 \times 507s = 2028s$. The images achieve a limiting point-source depth of AB(814W) = 27.2. 
For the strong lensing analysis, we focus on three strong lensing systems from the catalogue of lensing candidates \citep{faure}. All three are confirmed lenses and classified as \emph{best lenses} in \citep{faure}. 
For each system we can give a photometric redshift and the number of galaxies which are located within a circle of 1 arcmin radius. This measure indicates the density of the near neighbourhood and we can estimate the bias in the strong lensing shear measurement. 

\begin{table}[htp]
\caption{Properties of the strong lenses and their environment used for this work. We list the name of the lens and its photometric redshift $z_l$ as estimated by \cite{faure}. Further, we give the projected number of galaxies within a circle of radius 1 arcmin.} \
\begin{center}
\begin{tabular}{c|c|c}
strong lens system& $z_l$ & $N_{1'}$\\ \hline
COSMOS 0038+4133 & 0.89 & 130  \\
COSMOS 5921+0638 & 0.45 & 90  \\
COSMOS 5758+1525 & 0.38 & 99 
\end{tabular}
\end{center}
\label{tab:sl-env}
\end{table}
Especially the lens \emph{COSMOS0038+4133} is located in an over-dense region, although it is not part of a galaxy cluster \citep{faure2}. Compared to other lenses, the lens is located at a relatively high redshift, which explains the large value for the density measure.

\section{Strong lensing analysis}
\label{sec:sl-analysis}
The presence of image multiplicities in the strong lensing regime allows to determine aspects of both non-linear and linear lensing effects. In particular, line of sight shear has particular imprints in the observables such that a detailed modeling of the data allows to extract precise information on cosmic shear on the scale of the Einstein ring. The method to work requires a sufficient modeling on the HST pixel level and the simultaneous modeling of the lensing effect of the main deflector and external shear distortions as well as the surface brightness distribution of the deflector and source light. In Section \ref{sublesc:sl-basis} we present the details in our modeling procedure and in Section \ref{subsec:sl-results} we present the inference results from the modeling of the individual lenses.

\subsection{Strong lens modeling}
\label{sublesc:sl-basis}
For the analysis of three strong gravitational lensing systems we use \textsc{Lenstronomy}, a method and software\footnote{\url{https://github.com/sibirrer/lenstronomy}} presented in \citep{lenstronomy, birrer-basis}. We simultaneously forward model the lensing effect, the source surface brightness and the deflector light surface brightness and perform the model parameter posterior inference with an MCMC sampling.
For the lens mass distribution we choose a softened power-law elliptical mass distribution \citep[SPEMD,][]{barkana} and the line-of-sight lensing contribution we model as a linear shear component.
For each strong lens system we make an individual choice of basis sets for the surface brightness of the source  and lens galaxy and thus the reconstruction of their intrinsic surface brightness. 
For the deflector light we use a superposition of one or two elliptical S\'ersic models \citep{sersic} centered at the same position. For the lensed source galaxy, we use on top of an elliptical S\'ersic profile a set of shapelets \citep{shapelets1}, where the number of shapelet basis functions is chosen individually for each system: We use no shapelets to model the source of COSMOS5921+0638, while we use 3 (5) shapelet polynomial orders, $n_{\rm max}$, in order to model the sources of COSMOS0038+4133 (COSMOS5758+1525). \\
In addition, a uniform light component is added to the models to account for inaccuracies in the background subtraction of the data product. 
A key component in the modeling is an accurate Point Spread Function (PSF)
A SExtractor-analysis of the whole image is made to find stars nearby, similar to a TinyTim-simulation \citep{tinytim}, for a realistic representation of a PSF. We use a stacked image of stars with suitable magnitude and size in the near neighbourhood of the strong lens system as the PSF in our models.
The modeling procedure is adopted from \citep{birrer-basis} and Fig. 4 therein and the subsequent improvement by \cite{shajib2020}.
After having defined the parametrisation and accordingly priors, upper and lower bounds and a initial spread for the parameters of the models, a particle swarm optimisation (PSO) \citep{pso} is run. When this PSO is converged and we are satisfied with the resulting model the result of the PSO defines start parameters for a further investigation of the strong-lens-system model with a Monte-Carlo-Markov-Chain (MCMC). From the results of this MCMC we can draw conclusions for the physics of this strong lens system. Throughout the analysis, we chose flat priors in all the parameters we chose to infer.

\subsection{Analysing COSMOS lenses}
\label{subsec:sl-results}

\begin{figure}
\centering
\includegraphics[width=15cm]{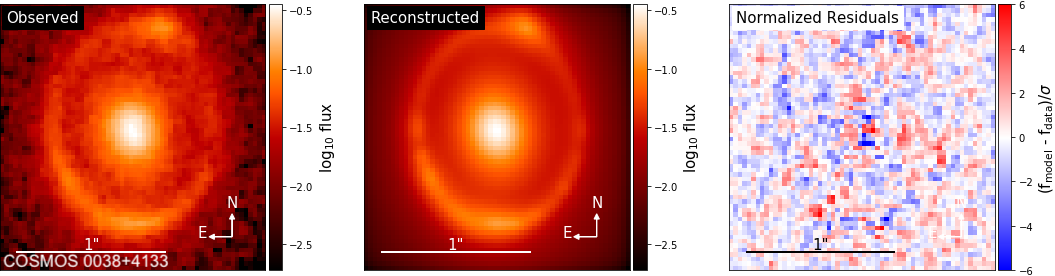}
\includegraphics[width=15cm]{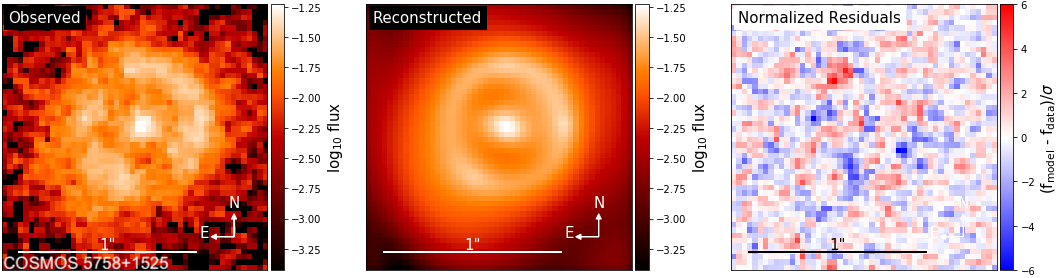}
\includegraphics[width=15cm]{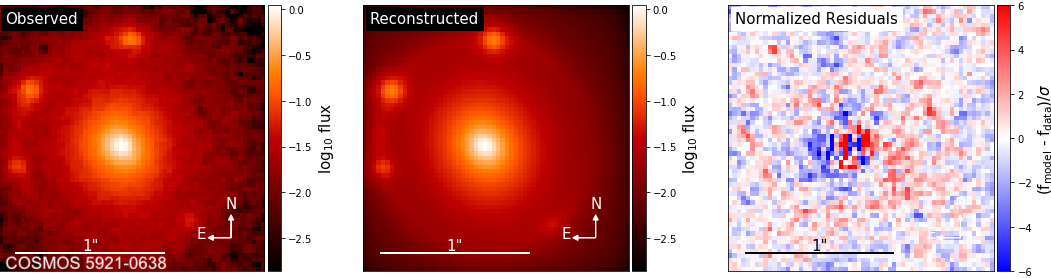}
\caption{Results of the strong lens reconstruction: In the left column there we show the cutout image of the strong lens system from the COSMOS survey, in the middle column we show the  model from the \textsc{lenstronomy}-reconstruction. The right panels show residuals. In all cases we were able to reconstruct the features of the lensing systems and to obtain the cosmic shear present in this region of the sky.}
\label{fig:sl-results}
\end{figure} 
The three lenses we are modeling are presented in Figure \ref{fig:sl-results}. Our goal is to reliably infer external shear components in the model that are not accounted by our deflector model choice. The extent of the extended Einstein ring is small and may not allow for a precise inference of the power-law slope of the main deflector profile. For this study, we fix the power-law slope of the lensing system $\gamma = 2$ to be isothermal. Extensions and their impact on the overall interpretation of the inferred shear values may involve a hierarchical approach to lens modeling and is beyond this study.
We present the results of the fitting in Table \ref{tab:sl-results} and Figure \ref{fig:sl-results}.\\
\begin{table}[htp]
\caption{Fitting results for three strong lens systems in the COSMOS survey. The strong lens analysis described in Section \ref{sec:sl-analysis} includes weak lensing perturbers near the lens galaxy ($\gamma_i$) as well. Table \ref{tab:sl-results} shows the best fits obtained and describes the relevant parameters of the system: the coordinates of the strong lens, the Einstein radius $\theta_E$. }
\begin{center}
\begin{tabular}{c|c|c|c}
& 0038+4133 & 5758+1525 & 5921+0638 \\ \hline 
  RA &150.159500 &149.494440 &149.840690 \\  
  DEC & 2.6927269 & 2.2569796 &2.1106531 \\
  $\theta_E$ &0.6191 $\pm$ 0.0003 & 0.322 $\pm$ 0.002 &0.712 $\pm$ 0.001 \\   
  $\gamma_1$  &-0.190$\pm$ 0.001 &$-0.032 \substack{+0.008 \\ -0.007}$ & 0.0001 $\pm$ 0.0003\\
 $\gamma_2$ &-0.000 $\pm$ 0.001 &-0.029 $\pm$ 0.009 & 0.006 $\pm$ 0.003 \\
\end{tabular}
\end{center}
\label{tab:sl-results}
\end{table}
We present in Fig. \ref{fig:sl-results} the result of the reconstructed strong lensing imaging data with our choices in the modeling. For all modeled systems we were able to reconstruct the main features of the strong lensing system, i.e. the rings with substructure in the first two systems and the four images in the last system as well as the light of the lensing galaxy to good quality allowing for a reliable cosmic shear estimate. In Figure \ref{fig:sl-results}, we show both the strong lens system as observed in the COSMOS survey (left column) as well as the image of the reconstructed strong lens system we obtained in our analysis (middle column). In the right column we show the normalised residuals, i.e. our way of comparing the reconstructed model with the original data. Only in the last system, the reconstruction of the lens light turned out to be very challenging: Even with a superposition of multiple Sersic profiles, we were not able to reconstruct the lens light perfectly. The spatial extent of the strong lens systems is described by the Einstein radius $\theta_E$, which we give alongside the shear components $\gamma_i$ characterising the influence of external sources along the line of sight in Table \ref{tab:sl-results}.\\

\section{Weak lensing analysis}\label{sec:wlanalysis}
\label{sec:wl-analysis}
\begin{figure}[h] \centering
\includegraphics[height=8cm]{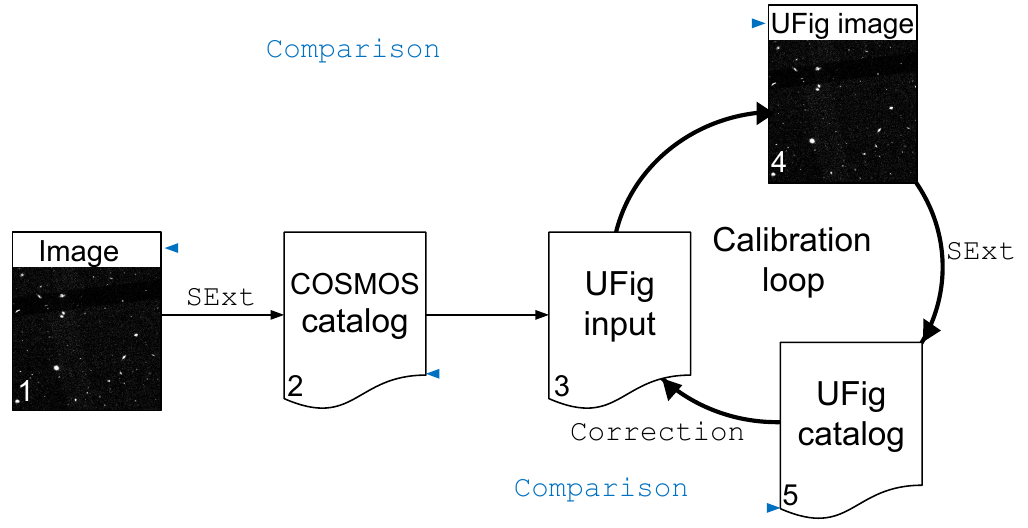}
\caption{Schematic diagram illustrating the cloning process described in Section \ref{sec:wlanalysis}.The original image from COSMOS (1) is analysed with SExtractor in order to obtain an object catalogue (2). Some information (e.g. object position and shape) in the catalogue is used as input for UFig (3) in order to create an image with UFig (4). This image is again analysed with SExtractor, the resulting catalogue (5) is corrected by comparison with (2) and then again used as UFig input (5 $\rightarrow 3$). Steps 3 to 5 are repeated until the first catalog (from the COSMOS image) is similar to the last one obtained from the last UFig simulated image.}
\label{fig:cloning}
\end{figure}
For the calibration of the shape measurement and the weak lensing analysis of the COSMOS images, we use \emph{cloning}, which employs techniques and tools used in \citep{bruderer, cb2016b, cb2018}. This method is schematically shown in Fig. \ref{fig:cloning}.
We first analyse the COSMOS image with SExtractor \citep{bertin} ($1 \rightarrow 2$). This analysis results in an initial catalog (2) containing detected objects (stars and galaxies) and their properties in the COSMOS image. Using the objects' measured properties from the catalog as an input ($2 \rightarrow 3$), we simulate a new image closely resembling the original COSMOS image ($3 \rightarrow 4$). This cloned image is then analysed with SExtractor as well ($4 \rightarrow 5$).
The original and cloned images and catalogues with the object properties are then compared by evaluating diagnostic distributions. From this comparison we draw conclusions on how to change input parameters (3) for a new UFig image in order to make the rendered UFig image (4) as similar as possible to the original COSMOS image (1). The last input object property catalog, which is the result of $n$ corrections from the cloning \emph{loop} is then used for the weak lensing analysis. \\
The different steps of the cloning scheme are described in more details in the following sections: In subsection \ref{subsec:sex} we describe the method to analyse the images and objects therein. In subsection \ref{subsec:ufig} we show how to create mock images from the previous analysis of original COSMOS images using UFig \citep{berge2013}. The optimisation of those simulations using the cloning scheme mentioned is described in subsection \ref{subsec:cloning} and the final shear analysis in the Cosmos field is shown in subsection \ref{subsec:shape}, where the shape measurement process is described. Finally, in \ref{subsec:wl-results} the measured shapes are related to the cosmic shear.
\subsection{Analysing COSMOS images with SExtractor}
\label{subsec:sex}
For our cloning method, we need a catalog of the identified objects and their properties in the COSMOS field for the further steps. For the creation of such a catalog we make use of SExtractor \citep{bertin}. This software analyses images, detects objects therein, and measures the properties of these objects. \\
We run SExtractor by employing the three-step \emph{hot-cold} scheme, as proposed by \citep{leauthaud}. In this method, SExtractor is run twice in order to  detect both bright and large, and faint and small objects. In a third step, the results are then combined while taking care of deblending effects (i.e. removing double counts).\\
Both samples are then merged and combined with a masking method to derive the final catalog. For further details on the SExtractor settings and the detection parameters for each step, we refer to \cite{leauthaud}.
\paragraph{Star-Galaxy-Separation}
In order to faithfully clone and reproduce the original images (section \ref{subsec:cloning}) and to select suitable galaxies for the lensing analysis (section \ref{subsec:shape}), we need to distinguish between stars and galaxies. For this step, we use a method proposed by \cite{leauthaud}.\\
We plot the measured brightness of the objects $\mu$(\texttt{MU\_MAX}) against their magnitude (\texttt{MAG\_AUTO}). The objects appearing on a straight line with $\mu \propto \textup{mag}$ are characterised as point sources (i.e. stars), the objects on a horizontal line at low $\mu$ are identified as saturated stars whereas the other objects are considered galaxies \citep[cf.][Fig. 5 therein]{leauthaud}. We select galaxies according to the following equation \ref{eq:sg-sep1}
\begin{align}
    \label{eq:sg-sep1}
    21 \leq \textup{MU\_MAX}:& & \textup{MU\_MAX} \geq \mu \cdot \textup{MAG\_AUTO} - 4.2 \\
    15 < \textup{MU\_MAX} < 21: & & \textup{MU\_MAX} > \mu \cdot \textup{MAG\_AUTO} -3.2 
\end{align}
with $\mu = 1.$
\\

\subsection{Creating images with UFig}
\label{subsec:ufig}
As described above, we clone the original COSMOS images by rendering new images with known inputs. For this step, we use UFig \citep[][]{berge2013, bruderer}, a program for the fast generation of images. Since UFig was mainly tested and developed for the simulation of images of ground-based quality (e.g. DES, \citep[e.g.][]{bruderer}), we had to develop a new software-plugin for the rendering of images with (space-based) COSMOS quality and COSMOS properties such as exposure time, magnitude zeropoint, number of exposures and other properties (see Section \ref{subsec:ufig} and Table \ref{tab:ufig-params}).\\
The code needs some general information on the COSMOS image (see below) as well as information for each single object to be rendered. UFig needs to know how many single exposure images have been stacked together for the COSMOS coadded image, the magnitude zeropoint, the gain, the exposure time, size and position in the sky, the rms of the background noise, the pixel scale, and the seeing (only used for the execution of SExtractor). These parameters are listed in the first set of parameters in Table \ref{tab:ufig-params}. \\
With UFig it is possible to render images of stars and galaxies with either randomly assigned properties given certain underlying models or by using object catalogs as input. Here, we use the latter and thus need to know the stars' and galaxies' positions, as well as their intrinsic magnitudes. For galaxies, we further need to know their intrinsic half-light-radius ($r_{50}$), have information on the shape of their light profiles, which we model with single-Sersic profiles \citep{sersic}, and their shapes (ellipticities $e_1,e_2$). These parameters are listed in the third and fourth sets of parameters in Table \ref{tab:ufig-params}.  \\
UFig furthermore allows the input of variable PSF maps. We therefore analyse non-saturated stars on the whole COSMOS field and create PSF maps using \textsc{HealPix}\citep{healpix} with a NSIDE value of 1024. These PSF maps contain information on size and shape of the PSF at a particular position in the sky, as this information varies across the imaging field. We model the PSF using a constant single-Moffat profile \citep{moffat1969}, where the shape parameter $\beta^{\textup{PSF}}$ is estimated on the data as well and fixed across the COSMOS field. The PSF parameters are listed in the second set of parameters in Table~\ref{tab:ufig-params}. 
\subsection{Optimising using the Cloning scheme}
\label{subsec:cloning}
In order to compare the reproduced, cloned images to the original ones, we have created a diagnostic toolbox. Using these diagnostics we can compare the distributions of the most important properties of objects measured on both the image and the catalog level. For the former, we compare the images pixel-by-pixel. For the latter, we first employ a matching method to find each objects counterpart, i.e. assigning the objects from the original COSMOS catalog an object from the SExtractor catalog of the newly rendered UFig image. This allows an object-for-object analysis, given that the cloned counterpart is detected (typically fewer objects are detected in the simulated images due to information loss). \\
To quantify the differences in the cloned images and catalogs, we use the metric
\begin{equation}
\xi_{f,g} = \frac{1}{N} \sum_i \left| f_i - g_i \right|.
\label{eq:xi}
\end{equation}
Here, $N$ denotes the number of bins in a histogram of the quantity in the catalog and image, $f_i$ and $g_i$ the values in bin $i$. For the image-level comparison, the pixels themselves are the bins. We have found this metric to be more robust than the sum of the squares due to the occurence of saturated pixels, difficult to model to a high accuracy with UFig.\\
While the position, brightness and PSF properties stay unchanged, we have to make adjustments for the shape parameters $r_{50}, e_1, e_2$ and Magnitude. These are adjusted with a linear correction function: $O_{f} = a*O_{i} + O_0$
In our analysis, we compare the $\xi$-value of the first cloned image (ie. without any corrections applied on the input catalog) to the last cloned image. By iteratively adjusting the corrections applied on the input catalogs, we can minimize this metric. \\
The final, adjusted input catalogs are then a measure of the "true", intrinsic object properties.\\
\subsection{Weak lensing analysis: Shape measurement}
\label{subsec:shape} 
We use the output catalogs computed in subsection \ref{subsec:cloning} to calibrate the shear measurement on these objects with another cloning loop and finally performing the measurement. In this calibration loop, we artificially add small lensing distortions to the galaxies to be rendered. By then comparing the shear estimated on the images to the artificial input shear, we can calibrate our shear analysis accordingly. This calibration scheme is described in this section. \\
To measure the objects' shapes, we follow the prescriptions described in \citep{cb2018}, which are themselves based on \citep[cf.][]{rhodes2000}. We use the weighted quadrupole moments as estimated by SExtractor $I_{11} := \texttt{X2WIN\_IMAGE,}$ $I_{12}:= \texttt{Y2WIN\_IMAGE}$ and $I_{22}:= \texttt{XYWIN\_IMAGE}$. We then perform an effective PSF-deconvolution by subtracting a multiple $\alpha$ of the PSF shape moments $P_{ij}$ from the measured total moments $I_{ij}$
\begin{equation}
Q_{ij} = I_{ij} - \alpha P_{ij}.
\label{eq:psf-corr}
\end{equation}
We then compute estimate the radius within which 50\% of a galaxies flux is contained (half-light-radius), $r_{50}$ and the ellipticities $e_i$ by combining the PSF-corrected quadrupole moments
\begin{equation}
\label{eq:shape}
    \begin{aligned}
    e_1 &= e_1(\alpha, \eta) =  \eta \frac{Q_{11}-Q_{22}}{Q_{11}+Q_{22}}, \\
    e_2 &= e_2(\alpha, \eta) =   \eta \frac{2Q_{12}}{Q_{11}+Q_{22}}, \\
    r_{50} &=  r_{50}(\alpha) =   \sqrt{2 \log(2) (Q_{11}+Q_{22})}.
    \end{aligned}
\end{equation}
For the lensing analysis presented in the following subsections, we thus first perform the star-galaxy-separation described by eq. \ref{eq:sg-sep1}. In order then to remove all objects too small and/or too faint for the shear analysis, we apply the following cuts
\begin{equation}
\label{eq:lensingcuts}
    \begin{aligned}
    &r_{50} > 1.6 \cdot r_{50}\left(P_{ij}\right) \\
    &\texttt{FLUX\_AUTO/FLUXERR\_AUTO} > 15.
    \end{aligned}
\end{equation}
Here, the size cut is relative to the local PSF size $r_{50}\left(P_{ij}\right)$. The brightness cut on the other hand is applied by evaluating the signal-to-noise ratio $\texttt{FLUX\_AUTO/FLUXERR\_AUTO}$, which needs to be larger than a certain threshold.
We calibrate both the PSF correction parameter $\alpha$ and the linear correction parameter $\eta$ on the simulated images by comparing the estimated sizes and ellipticities with the input values (cf. Fig. \ref{fig:cloning}). \\
In order then to use eq. \ref{eq:shear-susc} to estimate the galaxies' shear components, we need to estimate the shear susceptibility $G^\gamma$. We thus add another calibration loop and calibrate the relation. As described above, given the fiducial configuration output of our cloning scheme, we apply an artificial, small shear to the simulated galaxies and generate random realizations. For each shear component independently, we sample Gaussian distributions $\mathcal{N}(0,0.02)$ and assign these to the galaxies. The new images including shear, are re-analysed using the same measurement prescriptions. We can then to first order fit a linear relation between the input shear and the estimated shape. In fact, for reasons of numerical stability, we invert the input shear and the estimated shape and estimate the slope of the relation, which corresponds to the inverse susceptibility. Hence, the slopes of these relations are $\left(G^\gamma_i\right)^{-1}$, i.e.
\begin{equation}\label{eq:shear-est}
    \left\langle e_{out,i} \right\rangle = \left(G^\gamma_i\right)^{-1}\cdot\left\langle \gamma_{in,i} \right\rangle.
\end{equation}

\subsection{Weak lensing analysis: Diagnostics} 
\label{subsec:wl-results}
\begin{table}
\caption{Simulation parameters used for generating COSMOS-like images with UFig. The first set of parameters describes parameters fixed across the image derived from the COSMOS general image properties and constant parameters estimated on the data itself. Then, the PSF parameters are listed. These are estimated directly on the COSMOS data and the resulting variable PSF maps are used as input to the simulations. The third set of parameters are galaxy and star properties as estimated by SExtractor and used in the image generation without any corrections applied. Lastly, the fourth set of parameters are parameters galaxy and star properties as estimated by SExtractor, but adjusted using the calibration loops in the cloning scheme.} \centering
\label{tab:ufig-params}
\begin{tabular}{l|c|l}
name & value & description \\ \hline
\texttt{n\_exp} & 4 & Number of exposures   \\ 
\texttt{magzero} & 32.6995 & magnitude zeropoint \\
\texttt{gain} &1.0 & gain \\
\texttt{exposure\_time} & 507 & exposuire time in sec\\
\texttt{background\_sigma} & 2.1 & rms of the background noise \\
\texttt{seeing} & 0.09699 & PSF seeing (only for SExtractor analysis) \\
\texttt{pixscale} & 0.03 & arcsec per pixel \\
\texttt{ra0} & image property & right ascension zeropoint\\
\texttt{dec0} & image property & declination zeropoint \\ \hline
\texttt{psf\_type} & const. Moffat & PSF model \\
$\beta^{PSF}$ &2.07238 & PSF beta (fixed, estimated on data) \\
\texttt{psf\_e1, psf\_e2} &$e_{i}(P_{ij})$ & PSF ellipticity components (cf. eq. \ref{eq:shape}) \\ 
\texttt{psf\_r50} &$r_{50}(P_{ij})$ & PSF half light radius (cf. eq. \ref{eq:shape}) \\ \hline

\multirow{2}{*}{\texttt{mu}} & \multirow{2}{*}{\texttt{MU\_MAX}} & object brightness (used for \\
 & & star-galaxy-separation c.f. eq. \ref{eq:sg-sep1}) \\
\texttt{x} & \texttt{XWIN\_IMAGE} & object pixel position in x direction \\ 
\texttt{y} & \texttt{YWIN\_IMAGE} &object pixel position in y direction \\ \hline

\texttt{e1, e2} & $e_{i}(Q_{ij})$ & galaxy ellipticity components  (cf. eq. \ref{eq:shape})  \\ 
\texttt{r50} & $r_{50}(Q_{ij})$ & galaxy half light radius  (cf. eq. \ref{eq:shape}) \\
\texttt{mag} & \texttt{MAG\_AUTO} & object magnitude \\ 
\end{tabular}
\end{table}
\begin{figure}
    \centering
    \includegraphics[width=10cm]{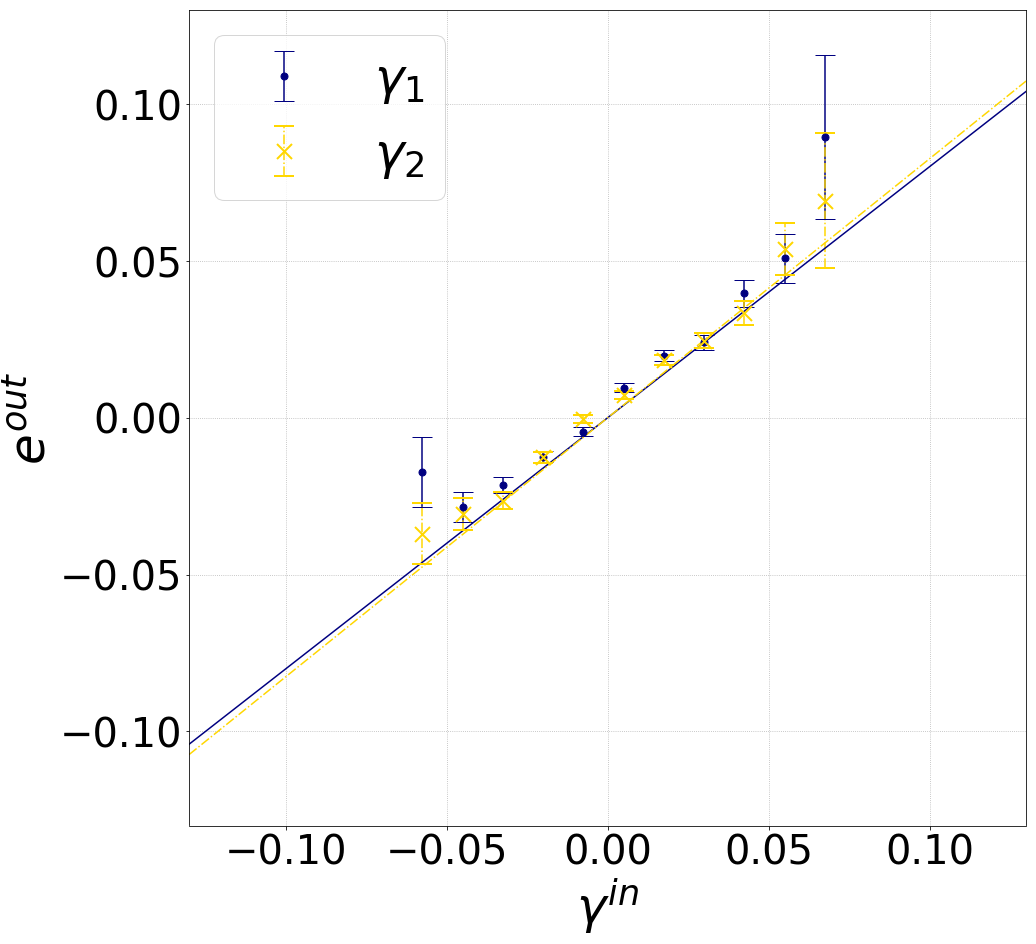}
    \caption{Fitted linear relations relating the ellipticity as estimated on simulations (y-axis) with the input shear (x-axis) (cf. Eq.~\ref{eq:shear-est}). The process of relating measured galaxy shapes to gravitational shear is described in \ref{subsec:wl-results}: By artificially adding a signal of shear $\gamma^{in}$ to the  galaxies simulated with UFig and then measuring the shear in the simulated image by averaging over galaxy ellipticities, we can calibrate the shear measurement and find the relation shown in eq. \ref{eq:shear-est-value}.}
    \label{fig:wl-shearcalibration}
\end{figure}
With the full cloning method, we are able to calibrate the shear measurement using the scheme described in section~\ref{subsec:shape}. The resulting relations between the estimated shape and the input shear used to estimate the shear susceptibility are shown in Fig. \ref{fig:wl-shearcalibration}. The fit is performed using the whole data set of $\sim 10^5$ galaxies (the bins in Fig.~\ref{fig:wl-shearcalibration} are only for visualization purposes). The fit yield values of $(G^\gamma_i)^{-1}$ = 0.80, 0.83. The relations for the two components thus agree well with each other. For the purpose of this work, i.e. to compare shear as estimated by weak and strong lensing analyses, we set a single value $G^{\gamma} = 1.25$ for the two components.
To first order, the relation between the estimated shape on the COSMOS galaxies and estimated shear is thus
\begin{equation}\label{eq:shear-est-value}
    \left\langle \gamma \right\rangle \sim 1.25 \left\langle e \right\rangle.
\end{equation}
As a qualitative, diagnostic test for our shear measurement, we use the method proposed by \cite{kaiser} to compute a map of the weak lensing convergence, which is a measure of the mass distribution. Our convergence map of the COSMOS field is shown in Fig. \ref{fig:massmap}. The brightness peaks indicate mass overdensities in the field. We find that these structures identified within this map agree well at the visual level with the structures in the map identified by \citep[e.g.][Fig 1]{massey}. 
\begin{figure} \centering
\includegraphics[width=10cm]{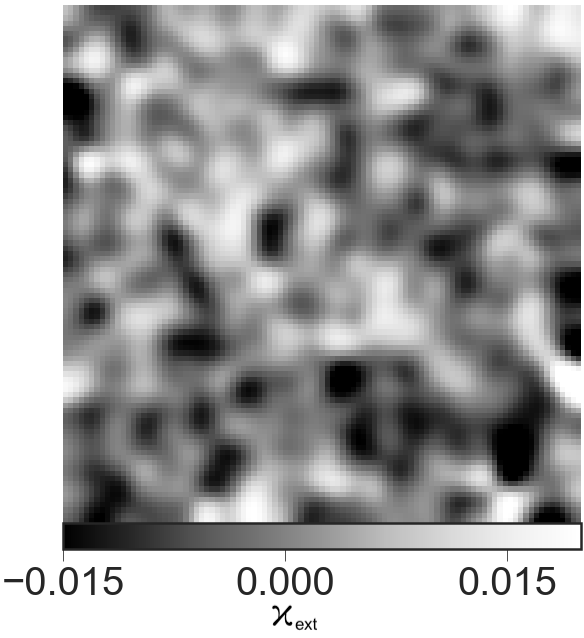}
\caption{Estimated convergence map of the COSMOS field. Bright spots indicate mass overdensities, whereas dark spots indicate voids in the cosmic web. We calculated the convergence $\varkappa$ with the method presented in \citep{kaiser} from our measured shear values from 332762 galaxies.
}
\label{fig:massmap}
\end{figure}
\begin{figure}[h]
\centering
\includegraphics[width=15cm]{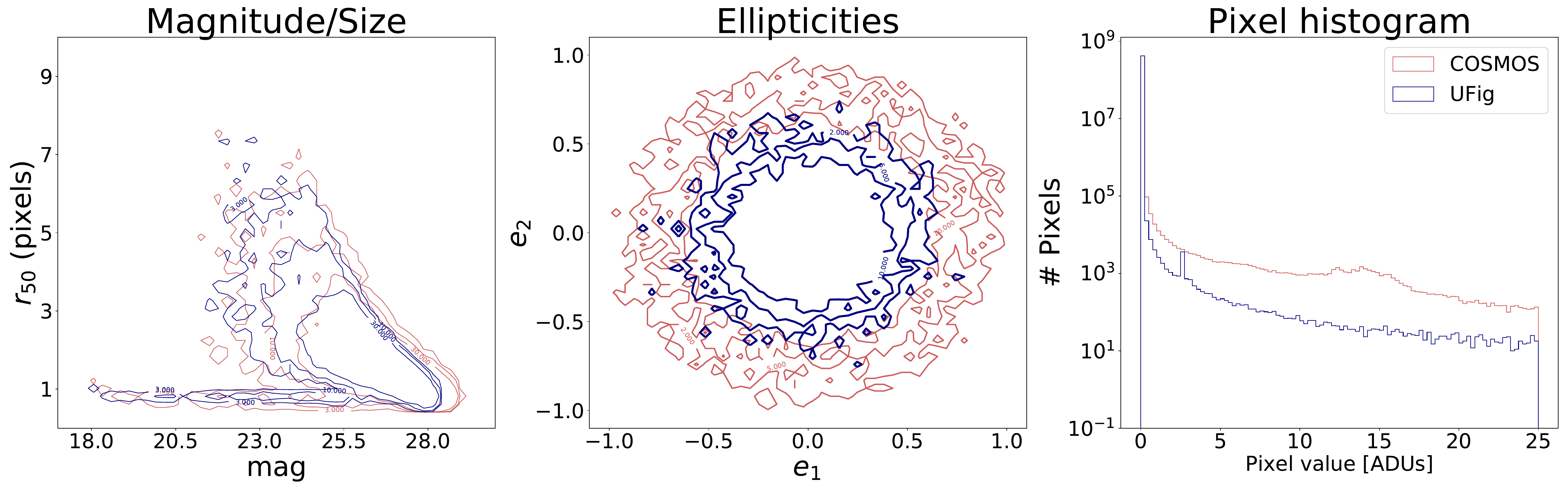}
\caption{Distributions used as diagnostics for the cloning process. The objects of both the COSMOS and the mock image are compared in the magnitude-size plane (left panel) and the ellipticity-1-2-plane (middle panel). For all distributions all stars and galaxies are considered, including the ones not used for the shear estimate. The right panel shows a histogram of the images' absolute pixel values.}
\label{fig:diagnostics}
\end{figure}
\begin{figure}
\centering
\includegraphics[width=15cm]{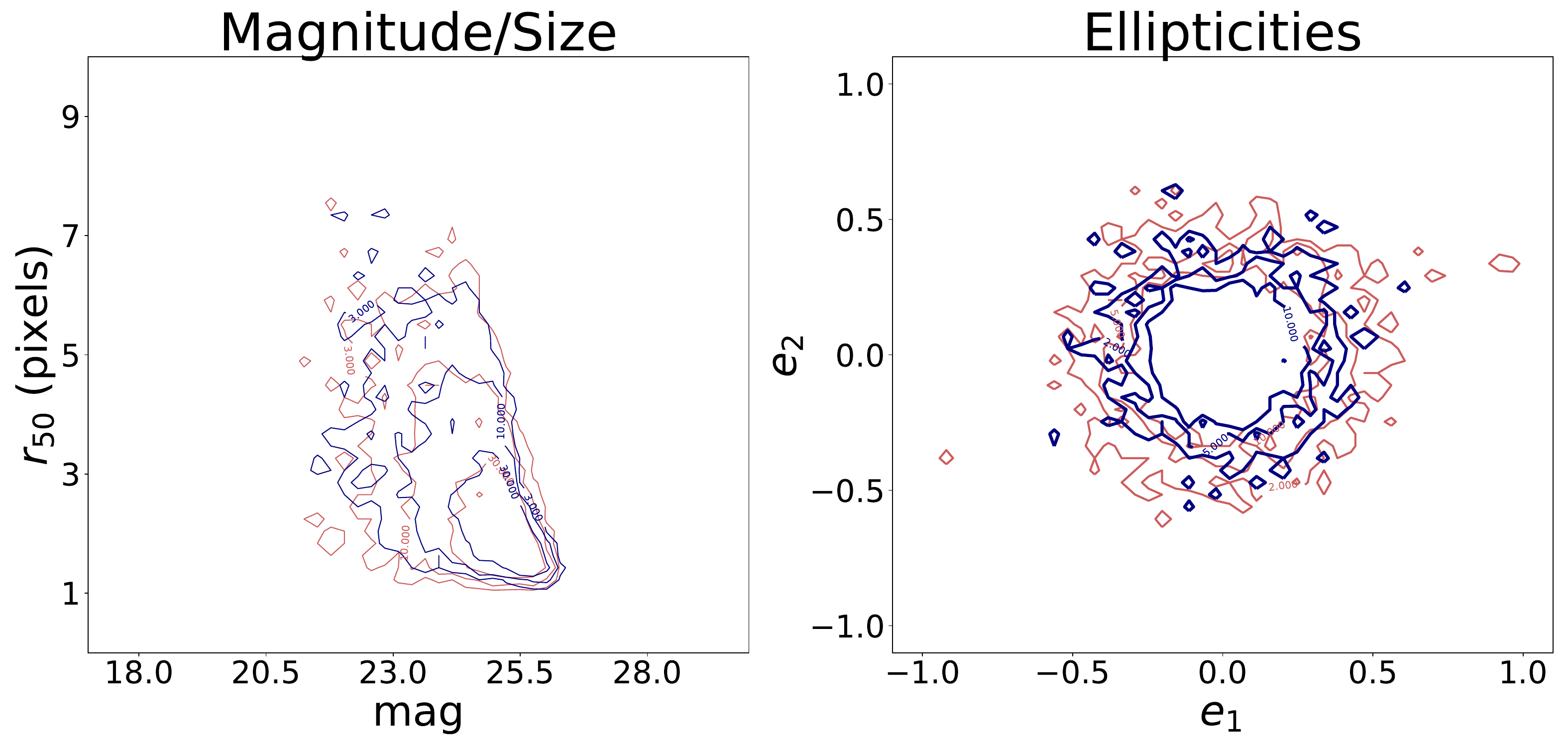}
\caption{Diagnostic distributions similar as in Fig. \ref{fig:diagnostics}, but only selecting the objects passing the lensing cuts. The selected galaxies of both the COSMOS and the mock image are measured and compared in the magnitude-size-plane (left panel) and the ellipticity-1-2-plane (right panel). Accurate shear calibrations based on simulations require a sufficient resemblance between the simulation and the sky.}
\label{fig:diagnostics2}
\end{figure}
As additional diagnostics, we perform analyses at the image- and the catalog-level. For the first comparison, we use the pixel values for the estimated and simulated images. At the catalog-level, we compare quantities as estimated by SExtractor.
Parts of this comparison are shown in Fig. \ref{fig:diagnostics}, where we show the distributions of five key quantities from one example image. The left panel shows the relation of size $r_{50}$ and magnitude $mag$ (\texttt{MAG\_AUTO}) for all objects, including stars. In the middle panel, the distribution of ellipticities in the $e_1, e_2$-plane is shown. These parameters were estimated using the method described in Section~\ref{subsec:shape}. Lastly, the right panel shows the number of counts per pixel in the COSMOS and UFig images.
While both the magnitude-size plane and the absolute pixel value histogram display a good agreement, the ellipticity contours are discrepant. This is explained by the reduced number of objects in the UFig image, which is also apparent in the other two panels. Since objects with a high ellipticity tend to be faint and small, these objects tend to not get identified by SExtractor when rendered again with UFig. This can be seen in Fig. \ref{fig:diagnostics2} where we show a comparison of the object parameters, only considering galaxies selected for the weak lensing analysis. Galaxies large and bright enough and used in the shear analysis after the lensing cuts are reproduced well by our cloning method.\\
For a more quantitative comparison, we use the error measure given in eq. \ref{eq:xi}. We can thus analyse the quality of the last cloned image and the first cloned image, which was rendered using the uncorrected catalog of objects estimated by SExtractor on the COSMOS images, with respect with the original image. For the three key quantities \texttt{MAG\_AUTO}, \texttt{FLUX\_RADIUS}, and \texttt{MU\_MAX} the values are shown in Tab. \ref{tab:cloning-error}. These values are reassuring, showing a clear improvement of our cloned images through our calibration scheme. The residual errors are mainly due to the loss of faint and small objects just above the detection threshold when re-rendering the image. Since these faint objects are removed by our weak lensing cuts (cf. Section \ref{subsec:shape}), not being able to detect then in the cloned image does not affect our weak lensing measurement. 
\begin{table}
\centering
\caption{Values of the $\xi$-metric (Eq.~\ref{eq:xi}) for quantities estimated on the real COSMOS and simulated UFig images. Here, the values for the estimated magnitude, size and surface brightness are listed. In the first column, the last cloned image, as output of the full cloning scheme, is used in the analysis. For the values in the second column, the first cloned image, which uses the uncalibrated object catalog as input, is used.}
\begin{tabular}{|c|c|c|}
\hline 
quantity & $\xi_{f,g}^{post}$ & $\xi_{f,g}^{pre}$ \\ 
\hline 
\texttt{MAG\_AUTO} & 54.69  & 84.552\\ 
\hline 
\texttt{FLUX\_RADIUS} & 54.32 & 83.174\\ 
\hline 
\texttt{MU\_MAX} & 54.69 & 86.866 \\ 
\hline 
\end{tabular} 
\label{tab:cloning-error}
\end{table}

\section{Joint analysis}
\label{sec:joint-analysis}


We performed a strong lens analysis using the basis sets by \citep{lenstronomy} for the three strong lens systems in order to infer the cosmic shear at the systems' specific positions (cf. Section \ref{sec:sl-analysis}). In parallel, we analysed the whole COSMOS field using a novel weak-lensing-analysis technique called cloning \citep[][]{bruderer,cb2016b,cb2018}, allowing us to estimate cosmic shear at every position in the field (cf. Section \ref{sec:wl-analysis}).
\begin{figure}
    \centering
   \includegraphics[width=15cm]{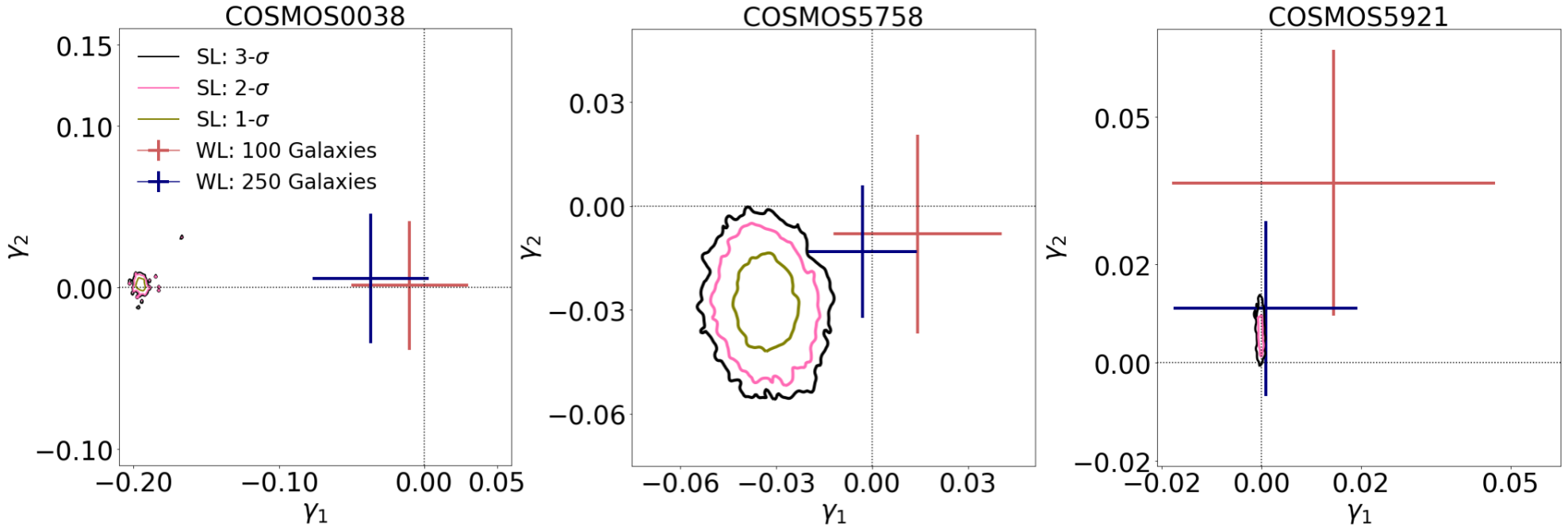}
    \caption{Comparison plot of strong and weak lensing cosmic shear estimates for three strong lens systems. For each strong lensing estimate we show the $1-\sigma$ (green), $2-\sigma$ (pink), and $3-\sigma$ (black) confidence level contours. For the weak lensing estimates, for which shear estimates are obtained by a calibrated average over the shapes of the 100 and 250 nearest galaxies, we show the $1-\sigma$ errors in the means. {While the left panel, representing the strong lens system COSMOS0038+4133 with the strong and weak lensing measurements in statistical tension in respect to the expected covariance (\ref{eq:covariance-matrix}) between both measurements, a potential sign for unaccounted systematics in either one or both of the measurement techniques; the middle and right panel (COSMOS5758+1525 and COSMOS5921+0638) show measurement results with deviations well within the expected covariance range calculated in eq. \ref{eq:covariance-matrix}. The tendency of the shear measurement to be degenerate with the measurement of the lens ellipticity shows the most clearly in the measurement of COSMOS5921+0638. For this lensing system, the strongly elongated shape of the strong lensing measurement in the $\gamma_2$-direction is caused by a strong degeneracy with $e_2^{lens}$ in the MCMC process.}}
    \label{fig:comparison}
\end{figure}
As a first step towards a joint strong and weak lensing analysis, we show weak lensing measurements at the position of the strong lensing line of sight and compare them with the independent strong lensing measurements in Fig. \ref{fig:comparison}. We chose two different weak lensing estimates, one by taking the average over the measured galaxy shapes of the nearest 100 galaxies, and one with averaging over 250 galaxies. The galaxies are chosen for both estimates within the redshift range of $z_l - 0.2 < z < z_l + 0.2$ where $z_l$ denotes the redshift of the strong lens deflector. For the weak lensing measurement, we show the errors of the mean derived from the shape noise uncertainties. We display the statistical errors of the strong lensing measurement as computed by the MCMC.\\
For two of the three systems, the absolute value of the weak lensing shear estimates tends to be smaller. Especially for COSMOS0038+4133, the strong lensing shear estimate is large and estimated to high accuracy, while the absolute value of the weak lensing estimates is small.\\
A difference of strong and weak lensing measurement is to be expected by  consideration of different length scales being averaged over in the two measurements. While for the strong lens measurement only the small region of the strong lens itself is considered ($\sim 1$ arcsec$^2$), the weak lensing measurement makes use of several hundred galaxies distributed over a region up to a few arcminutes$^2$.
For an estimate of this expected difference, we compute the anticipated covariance between this strong and weak lensing measurement using eq. \ref{eq:shearvar} for the diagonal elements and eq. \ref{eq:cls-offdiag} for the off-diagonal elements. For this computation we use the \textsc{PyCosmo} package \citep{refregier2018,tarsitano2020}. Here, we aim to provide a qualitative measure of the discrepancy. We hence set $f_{d,s}=1$ for the weight function for this order of magnitude estimate. We also assume a fixed cosmology with $h=0.7, \Omega_m = 0.3, \Omega_b = 0.06, \sigma_8 = 0.8$. We include the redshift distribution of the three strong lens system surroundings by using redshift data by \citep{laigle}.\\
Our calculation yields the covariance matrix
\begin{equation}
\label{eq:covariance-matrix}
\textup{cov(SL,WL)}=
    \begin{pmatrix}
 1.6 \times 10^{-3} & 9 \times 10^{-4}\\ 
9 \times 10^{-4}& 4 \times 10^{-4}
\end{pmatrix}.
\end{equation}
Here, the first entry in the upper left corner denotes the SL-variance, the one in the lower right corner the WL-variance. The off-diagonal elements denote the SL-WL-covariance. This value for the covariance tends to be smaller than the difference observed for the individual strong lens systems in Fig. \ref{fig:comparison}. \\
{This indicates that more data is required for a signal. In order to estimate the number of strong lens systems needed to measure the strength of the effect for an effective cross-calibration,} we sample n tuples from a multivariate normal distribution characterised by the covariance matrix calculated above. We then compute the variance of 500 different draws for each value $\left\langle \textup{SL}, \textup{SL} \right\rangle, \left\langle \textup{WL}, \textup{WL} \right\rangle$, and $\left\langle \textup{SL}, \textup{WL} \right\rangle$. As this value decreases with $n^{-1/2}$, we find that one needs 48 (17) strong lens systems to get to a $5 \sigma ~(3 \sigma)$ level of confidence in the estimation of the covariance of both measurements. \\
{It is however possible to detect the cross-correlation signal with the given data. We then analyse the covariance between strong and weak lensing and find a $2 \sigma$-detection of the cross-correlation signal when using the nearest 200 galaxies for the weak lensing analysis. When considering only the 100 closest neighbouring galaxies, the significance of the detection of the cross-correlation signal decreases to $1-\sigma$, which we mostly attributed to the lower signal-to-noise in the weak lensing measurement. Our fiducial measurement in this work is using 200 galaxies in the weak lensing analysis.}
The effect of averaging over different scales however does not allow us to fully explain the discrepant estimates for the strong lens system COSMOS0038+4133, where the difference is much larger than the predicted strong-weak-lensing-covariance. This system has been analysed before by \citep{birrer-welschen}, who have also found large values for the cosmic shear. It is therefore likely that in this region there is a strong nearby source of cosmic shear induced on the strong lensing estimate, but on too small scales to significantly affect the weak lensing shear estimate. To improve the strong lensing shear analysis, one might also have to include nonlinear shear in the strong lens modelling arriving from multi-plane effects coupling the line of sight structure with the main deflector. Although there was no cluster detected nearby \citep{faure2} the high density of galaxies along the line of sight might influence the local shear at the strong lens system, which is too small to resolve with the weak lensing analysis. Another hypothesis is that there is an effective shear term induced by the assumptions on the radial extend of the deflector mass model leading to artificial shear terms (see \citep{vandev2020}). \\
The second system, COSMOS5758+1525 shows only a small discrepancy between the strong and weak lensing shear measurement. The remaining difference might be explained by the poor signal to noise ratio in the data for this particular system (cf. Fig. \ref{fig:sl-results}, left middle panel). \\
For the third system, COSMOS5921+0638 both measurements are in good agreement. The data has an excellent signal to noise ratio, shows clear strong lensing features (i.e. four point sources and a faint arc connecting them) and could be reproduced in the modelling process. The background density in the neighbourhood of the strong lens system was fairly low (as it was also the case for the second system), and it hence yields similar shear estimates as the weak lensing measurements. The elongation of the strong lensing shear estimate in $\gamma_2$-direction as shown in Fig. \ref{fig:comparison} shows a degeneracy of the shear estimate with the intrinsic lens ellipticity.\\
This joint analysis potentially allows for two different ways of combining shear measurements. On one hand, given multiple reliable strong lens measurements without nearby large overdensities (e.g. COSMOS5758+1525 or COSMOS5921+0638), it is possible to calibrate weak lensing analyses with strong lensing estimates. \\
On the other hand, the strong lensing analysis is influenced by biases or systematic measurement errors as well and may benefit by including the weak lensing shear estimates. For instance, in the analysis of COSMOS+0038, it is possible that the predominant local matter distribution strongly influences the  lens estimate of the shear. As the shear inference in the strong lensing analysis relies on an accurate representation of the main deflector profile, a missmatch in the modeling of the larger extents of the main lensing deflector can thus lead to effective shear terms misinterpreted for cosmic shear.

\section{Conclusion}
\label{sec:discussion}
In this paper, we have performed a first joint strong and weak lensing measurement of cosmic shear over an extended area on the sky of the cosmos field \citep{cosmos}.
To achieve independent measurements of cosmic shear from strong and weak lensing, we have forward modeling the effect of cosmic shear on three strong lensing systems. We have performed this measurement by separating the mass density profile of the main deflector and the linear shear components using the software \textsc{lenstronomy} \citep{lenstronomy}. This has yielded precise estimates of the linear shear components at the positions of the strong lens systems on scales of their Einstein rings \citep{birrer-basis}.
Independently, we have measured the weak lensing signal and estimated the shear field over the full COSMOS area with a novel method called \emph{cloning}. We have created a framework designed to reproduce the real COSMOS images using image generations by the Ultra Fast Image Gimulator \textsc{UFig} \citep{berge2013}. This has allowed us to first calibrate the shear measurement method for a fiducial simulation configuration for COSMOS-quality images and then secondly apply the calibrated measurement method on the actual COSMOS images.
We have thus been able to measure with two complimentary probes cosmic shear along the line of sight at the positions of three strong gravitational lenses and their surroundings.
As strong and weak lensing measurements cover different redshift distributions and angular scales, we have computed the expected strong-weak-lensing cross correlation quantifying the covariance between the two probes. We have then compared the measurements of the strong and weak lensing analyses with this expected covariance. For one of the strong lens systems (COSMOS0038+4133), we have found a discrepancy between the strong and weak lensing estimates despite taking the different angular smoothing and redshift distribution kernel of the two methods into account. This deviation could be an indication of either or a combination of an unaccounted selection effect of the strong lensing system, local lensing perturbations, or a sign of a deviation of the theoretical prediction of the power-spectrum at small angular scales.
Taking these results at face value and evaluating the prospects in the near future, we find that about 50 strong lens systems are needed in order to reliably measure the strong-weak lensing covariance for data similar in quality to COSMOS { in order to perform a cross-calibration of the two measurements. It is however already possible to detect a $\sim 2 \sigma$ signal of cross-correlation itself between the two measurements with the three strong lenses analyzed in this work.} \\
Our work has shown the prospects for both, calibrating weak lensing analyses with strong lensing measurements, and to calibrate selection effects and local lensing properties for strong lensing analyses. \\
Our results are pathfinders to incorporate large scale weak lensing information in the modeling of strong gravitational lenses and vice versa. Especially with automated strong lens analyses of upcoming large surveys such as Rubin, Roman and Euclid Observatories, cross-validation and -calibration of measurement methods become essential to probe the dark sector of our Universe.

\section*{Acknowledgements}
FAK is fully funded by the Foundation of German Business (\textit{Stiftung der deutschen Wirtschaft}, \url{sdw.org}). We made use of the software packages 
lenstronomy \citep{lenstronomy}, cosmohammer \citep{cosmohammer}, SExtractor \citep{bertin}, HealPix \citep{healpix}, TinyTim \citep{tinytim}, Numpy, SciPy, AstroPy, and UFig \citep{berge2013}. 

\bibliographystyle{JHEP}
\bibliography{references}

\end{document}